\documentstyle[12pt,a4]{article}

\begin{document}

\begin{flushright} 
NIKHEF/99-026
\end{flushright}

\noindent
{\large{\bf Killing-Yano tensors, non-standard supersymmetries and 
an index theorem}} \hfill \newline

\noindent
{\bf J.W.\ van Holten} \newline 

\noindent 
Theoretical Physics Group, NIKHEF \newline 
P.O.\ Box 41882, 1009 DB Amsterdam NL \newline 
{\tt vholten@nikhef.nl} \newline 
 

\noindent
{\small
The existence of Killing-Yano tensors on space-times can be probed by 
spinning particles. Specifically, Dirac particles possess new fermionic 
constants of motion corresponding to non-standard supersymmetries on 
the particle worldline. A geometrical duality connects space-times with 
Killing-Yano structure, but without torsion, to other space-times with 
Killing-Yano structure and torsion. A relation between the indices of
the Dirac-operators on the dual space-times allows to express the 
index on the space-time with torsion in terms of that of the space-time 
without torsion. 
} \newline
\footnoterule 
~\hfill \newline

\noindent 
It is standard procedure in general relativity to introduce 
the notion of test particles, idealized as structureless 
infinitely small mass points, to probe the geometry of 
space-time by identifying their orbits with geodesics \cite{AE1}. 
However, not all the geometric properties of space-time are 
encoded in the geodesics. Specifically, there can be structures 
on the space-time related to rotation, which become manifest 
in the dynamics of spin. Examples of spin-related structures 
which can live on a space-time are torsion and Killing-Yano 
tensors. 

Representing the particle's worldine co-ordinates at proper 
time $\tau$ by $x^{\mu}(\tau)$, and the components of its 
spin by the anti-symmetric tensor of dipole moments 
$S^{\mu\nu}(\tau)$ (following ref.\cite{JWH1}), the motion 
of the spinning particle in a torsion-free space-time is 
described by a generalization of the geodesic equation 
including coupling of the spin to the background curvature 
\cite{AP}:
\begin{equation}
\begin{array}{lll} 
\displaystyle{ \frac{D^2 x^{\mu}}{D\tau^2} }& = & 
\displaystyle{ \frac{1}{2}\, S^{\kappa\lambda} 
  R_{\kappa\lambda\;\;\nu}^{\;\;\;\;\mu}\, \dot{x}^{\nu}, }\\
 & & \\ 
\displaystyle{ \frac{D S^{\mu\nu}}{D\tau} }& = & 0. 
\end{array} 
\label{e.1}
\end{equation} 
Thus the spin-tensor is covariantly constant. The covariant 
world-line derivative in these equations represents the 
pull-back of the Riemann-Christoffel connection to the 
world-line. The standard geodesic equation is reobtained 
in the limit $S^{\mu\nu} = 0$. 

\noindent
It is not straightforward in general to derive these equations 
from a variational principle. The way we proceed here is to 
represent the spin of the particle in terms of anti-commuting 
c-numbers, the Grassmann-odd co-ordinates $\psi^{\mu}(\tau)$, 
as:
\begin{equation} 
S^{\mu\nu} = - S^{\nu\mu} = -i \psi^{\mu} \psi^{\nu}. 
\label{e.2}
\end{equation} 
In this formalism the equations of motion (\ref{e.1}) are 
obtained directly from the supersymmetric extension of the 
action for geodesic motion \cite{JWH2,Khr}. The price to be 
paid is of course that the spin variable has no direct 
physical interpretation; this can be overcome by interpreting 
it as a symbol for the spin of particles in quantum theory
\cite{Fad}, the physical value of the spin-components being 
obtained by an averaging procedure over all spin-histories 
along the world-line. Alternatively one may proceed directly 
to a hamiltonian description. This we will also do, but we 
will keep the representation (\ref{e.2}) of the spin-components 
as it allows to uncover a remarkable rich structure of 
symmetries and conservation laws which are obtained only by 
more cumbersome guesswork otherwise. This guess work can be 
much simplified by first developing the theory in terms of 
Grassmann co-ordinates, writing down and solving the equations 
of motion, and finally omitting all aspects having to do 
with the Grassmannian construction of $S^{\mu\nu}$, if desired. 

In terms of the covariant momentum $\Pi_{\mu}$: 
\begin{equation} 
\Pi_{\mu} = p_{\mu} - \frac{1}{2}\, \omega_{\mu} \cdot S 
 = g_{\mu\nu} \dot{x}^{\nu},  
\label{e.4}
\end{equation} 
with $p_{\mu}$ the canonical momentum and $\omega_{\mu}$ the 
spin-connection, the Hamiltonian for the spinning particle is 
\begin{equation} 
H = \frac{1}{2}\, g^{\mu\nu} \Pi_{\mu} \Pi_{\nu}.
\label{e.3}
\end{equation} 
The evolution of any scalar function on the phase-space is 
described by the Poisson-Dirac bracket \cite{JWH3}
\begin{equation} 
\begin{array}{l} 
\displaystyle{ \frac{dA}{d\tau}\, =\, \left\{ A, H \right\}, 
 }\\ 
 \\ 
\displaystyle{ \left\{ A, B \right\}\, =\,  
  D_{\mu} A \frac{\partial B}{\partial \Pi_{\mu}}\, -\, 
  \frac{\partial A}{\partial \Pi_{\mu}} D_{\mu} B\, +\, 
  R_{\mu\nu} \frac{\partial A}{\partial \Pi_{\mu}} 
  \frac{\partial B}{\partial \Pi_{\mu}} + i(-1)^A 
  g^{\mu\nu} \frac{\partial A}{\partial \psi^{\mu}}
  \frac{\partial B}{\partial \psi^{\nu}},
} 
\end{array} 
\label{e.5}
\end{equation} 
with the covariant derivatives defined in \cite{JWH3}.
Constants of motion can be found by requiring scalar 
phase-space functions to commute with the Hamiltonian 
in the sense of the brackets (\ref{e.5}). A universal 
constant of motion for the theories discussed here is
the supercharge $Q$:
\begin{equation} 
Q\, =\,  \Pi_{\mu} \psi^{\nu}, \hspace{2em} 
 \left\{ Q, H \right\}\, =\, 0, \hspace{2em} 
 \left\{ Q, Q\right\}\, =\, -2iH. 
\label{e.6}
\end{equation} 
The physical interpretation of this equation is simple: 
$Q = 0$ is the condition for the time-components of the 
spin to vanish in the restframe. From (\ref{e.6}) this 
condition is now seen to be compatible with the dynamical 
equations. Other constants of motion exist if the space-time 
admits additional structures like Killing vectors and tensors. 
Generalizing the construction for spinless particles, in 
the presence of a Killing vector field $K_{\mu}(x)$ 
there is a constant 
\begin{equation} 
J(x,\Pi,\psi)\, =\, K^{\mu}\, \Pi_{\mu}\, +\, \frac{1}{2}\, 
  B_{\mu\nu}\, S^{\mu\nu}, 
\label{e.7}
\end{equation} 
with $2 B_{\mu\nu} = K_{\nu,\mu} - K_{\mu,\nu}$. In fact, 
this quantity is not only constant, but also superinvariant: 
$\left\{ J, Q \right\} = 0$. Similarly, the existence of a 
symmetric Killing tensor $K_{\mu\nu}$ implies another 
constant of motion 
\begin{equation} 
Z(x,\Pi,\psi)\, =\, \frac{1}{2}\, K^{\mu\nu} \Pi_{\mu} 
 \Pi_{\nu}\, -\, \frac{1}{2}\, S^{\mu\nu} I_{\mu\nu}^{\lambda}
 \Pi_{\lambda}\, -\, \frac{1}{4}\, S^{\mu\nu} S^{\kappa\lambda} 
 G_{\mu\nu\kappa\lambda}, 
\label{e.8}
\end{equation} 
with the tensors $I_{\mu\nu}^{\lambda}(x)$ and 
$G_{\mu\nu\kappa\lambda}(x)$ solutions of the differential 
equations 
\begin{equation} 
I_{\mu\nu(\kappa;\lambda)} =  R_{\mu\nu\sigma(\kappa} 
  K_{\lambda)}^{\;\;\sigma}, \hspace{2em} 
G_{\mu\nu\kappa\lambda;\rho} = R_{\sigma\rho[\mu\nu} 
I_{\kappa\lambda]} ^{\sigma}. 
\label{e.9}
\end{equation} 
Clearly, in the absence of spin one gets the usual constants 
of motion associated with Killing vectors and tensors. In 
the presence of spin there are additional terms reflecting 
the spin-orbit coupling. Moreover, there can also be constants 
of motion that exist {\em only} for spinning particles. We 
have already discussed the standard supercharge (\ref{e.6}); 
but additional conserved supercharges may exist if the 
background geometry admits a Killing-Yano tensor $f_{\mu\nu}$
\cite{JWH3}. In such a geometry there exists an additional 
superinvariant constant of motion $Q_f(x,\Pi,\psi)$ defined by
\begin{equation}
Q_f\, =\, f^{\;\;\nu}_{\mu} \Pi_{\nu} \psi^{\mu}\, +\, 
  \frac{i}{3!}\, H_{\mu\nu\lambda} \psi^{\mu} \psi^{\nu} 
  \psi^{\lambda}. 
\label{e.10}
\end{equation} 
Here $f^{\mu}_{\;\;\nu}$ are the mixed components of a 
Killing-Yano 2-form $f = f_{\mu\nu}\, dx^{\mu} \wedge dx^{\nu}$,
which by definition posesses a 3-form field strength $H = df$ 
with the property $H_{\mu\nu\lambda} = f_{[\mu\nu;\lambda]} = 
f_{\mu\nu;\lambda}$. The superinvariance implies $\left\{ Q_f, 
Q \right\} = 0$. The most interesting aspect of the bracket 
algebra for the new supercharge is the relation 
\begin{equation} 
\left\{ Q_f, Q_f \right\}\, =\, -2iZ, 
\label{e.11}
\end{equation} 
with $Z$ of the form (\ref{e.8}), and  
\begin{equation} 
\begin{array}{l} 
K_{\mu\nu} = f_{\mu\lambda} f_{\nu}^{\;\;\lambda}, \hspace{3em}   
 G_{\mu\nu\lambda\kappa}  =  R_{\mu\nu\rho\sigma} 
 f_{\lambda}^{\;\;\rho} f_{\kappa}^{\;\;\sigma}\, +\, \frac{1}{2}\, 
 H_{\mu\nu}^{\;\;\;\;\sigma} H_{\lambda\kappa\sigma}, \\
 \\
I_{\mu\nu\lambda} = f_{\mu}^{\;\;\sigma} f_{\nu\lambda;\sigma}\, 
 -\, f_{\nu}^{\;\;\sigma} f_{\mu\lambda;\sigma} + H_{\mu\nu\sigma} 
 f_{\lambda}^{\;\;\sigma}. 
\end{array} 
\label{e.12} 
\end{equation} 
In the special case that $Z = H$ we have standard $N=2$ supersymmetry. 
In all cases with $H \neq 0$ this is impossible, and we have a true 
non-standard supersymmetry. Examples of such structures can be found 
e.g.\ in Kerr-Newman or Taub-NUT space-time \cite{JWH3,JWH4}. 
Killing-Yano tensors have been discussed previously in the 
context of non-standard Dirac operators which can be diagonalized 
simultaneously with the standard one \cite{CMcL,McLS}. These 
correspond precisely to the quantum-mechanical version of the 
supercharges presented above, in the sense of the correspondence 
relation 
\begin{equation}
p_{\mu}\, \rightarrow\, -i \partial_{\mu}, \hspace{2em} 
 \psi^{\mu}\, \rightarrow\, \frac{i}{\sqrt{2}}\, \gamma_5 
 \gamma^{\mu}, 
\label{e.13}
\end{equation} 
with $\gamma^{\mu}$ the local version of the Dirac matrices: 
$\left\{ \gamma^{\mu}, \gamma^{\nu} \right\} = 2 g^{\mu\nu}$. 
This  correspondence leads to the result that modulo a factor 
$\sqrt{2}$ 
\begin{equation} 
Q\, \rightarrow\, \gamma_5 \gamma^{\mu} D_{\mu} \equiv \gamma_5 
 {\cal D}, \hspace{2em} Q_f\, \rightarrow\, \gamma_5 \gamma^{\mu} 
 \left( f_{\mu}^{\;\;\nu} D_{\nu} - \frac{1}{3!}\, 
 \sigma^{\kappa\lambda} H_{\mu\kappa\lambda} \right) \equiv 
 \gamma_5 {\cal D}_{f}. 
\label{e.14}
\end{equation} 
Like the pseudo-classical supercharges these Dirac-operators 
anti-commute; defining ${\cal D}_f^5 \equiv \gamma_5 {\cal D}_f$, 
it then follows that ${\cal D}$ and ${\cal D}_f^5$ commute: 
$ [{\cal D}, {\cal D}_f^5]\, =\, 0 
$. This has some interesting consequences; in particular, the 
index of these two operators, defined as the difference between 
left- and righthanded zero-modes: $\Delta = n_+^0 - n_-^0$, is the 
same. In fact, only the simultaneous zero-modes of these 
operators can produce a non-vanishing contribution to the trace 
of $\gamma_5$ over all of the physical state space; hence \cite{HWP}: 
\begin{equation} 
\mbox{Tr}\, \gamma_5 = \Delta\,[{\cal D}] = \Delta\,[{\cal D}_f^5]. 
\label{e.16}
\end{equation} 
To evaluate such traces over infinite-dimensional spaces one has 
to regularize the expressions (\ref{e.16}). A convenient way is 
to do this in terms of the Witten-index of the corresponding 
supersymmetric quantum-mechanical model \cite{AG}:
\begin{equation} 
\Delta\, [{\cal D}]\, =\, \lim_{\beta \rightarrow 0}\,  
  \mbox{Tr}\, \left( (-1)^F e^{- \beta H} \right). 
\label{e.17}
\end{equation} 
This expression can be rewritten as a path-integral with the 
pseudo-classical action given by the hamiltonian (\ref{e.3}) 
of our supersymmetric spinning particle model, using periodic 
boundary conditions for the fermionic degrees of freedom 
$\psi^{\mu}$. Eq.(\ref{e.16})  now suggests one should also be 
able to write this quantity as a regularized trace 
\begin{equation} 
\Delta\, [{\cal D}_f^5]\, =\, \lim_{\beta \rightarrow 0}\,  
  \mbox{Tr}\, \left( (-1)^F e^{- \beta Z} \right). 
\label{e.18}
\end{equation} 
where we have replaced ${\cal D}$ by ${\cal D}_f^5$, and 
correspondingly $H$ by the square of the Killing-Yano 
supercharge $Z$. This corresponds to a theory in which the 
roles of Hamiltonian and Killing-tensor have been interchanged, 
as well as those of the supercharge $Q$ and the Killing-Yano 
supercharge $Q_f$. In ref.\cite{JWH5} this dual relation 
between metrics and Killing tensors, which has been observed 
independently in \cite{CF}, and between supercharges and  
Killing-Yano tensors was investigated systematically. It 
was shown, that for spinning particles the procedure in 
general works between the original space (without torsion), 
and the Killing-dual space only if the latter admits torsion. 
However, it is then no longer clear if the procedure (\ref{e.18}) 
to compute the trace of $\gamma_5$ has the same meaning as  
previously, as now it refers to the index of a Dirac operator 
on a different space-time, with different metric and with 
torsion, to which is generally added the problem of having to 
include boundary contributions \cite{PW}. On the other hand, 
if the equality still holds, the procedure can be turned around 
to express the index of a Dirac operator on some specific space-time 
with torsion in terms of that of another Dirac operator on a 
space-time without torsion. Recently Peeters and Waldron have 
managed to do the computation of the index on the Killing-dual 
space-times with torsion and non-empty boundary directly \cite{PW}; 
in the specific examples they have checked, they found it to agree 
with the known result for the original Dirac operator.

\end{document}